\documentclass[a4paper,11pt]{article}
\usepackage{pos, amsmath, bm}
\usepackage{graphicx}
\usepackage{multirow}

\title{Towards imaging Earth's large-scale structures by directional geoneutrino detection with Ocean Bottom Detector}

\author*[a]{Zhihao Xu}
\author[a]{Takumi Araki}
\author[a]{Simran Chauhan}
\author[b]{Brian C. Crow}
\author[b]{\\Max A. A. Dornfest}
\author[b]{Stephen T. Dye}
\author[b]{John Graham}
\author[a]{Misaki Hosoya}
\author[a]{\\Kunio Inoue}
\author[b]{John G. Learned}
\author[c]{Viacheslav A. Li}
\author[d, e]{William F. McDonough}
\author[a]{\\Takeru Ohno}
\author[a]{Takanobu Ono}
\author[a, \dagger]{Taichi Sakai}
\author[b]{Jackson Seligman}
\author[b]{\\Nathan Sibert}
\author[f]{David Vartanyan}
\author[a]{Hiroko Watanabe}
\author[b]{Jeffrey Yepez}
\onbehalf{\\(OBD Consortium)}

\affiliation[a]{Research Center for Neutrino Science, Tohoku University, 
  Sendai, Miyagi 980-8578, Japan}

\affiliation[b]{Department of Physics and Astronomy, University of Hawai'i at Mānoa, 
Honolulu, Hawaii 96822, USA}

\affiliation[c]{Lawrence Livermore National Laboratory, 
Livermore, California 94550, USA}

\affiliation[d]{Advanced Institute for Marine Ecosystem Change (WPI-AIMEC), Tohoku University \& JAMSTEC, 
Sendai, Miyagi 980-8578, Japan}

\affiliation[e]{Department of Geology, University of Maryland, College Park, 
Maryland 20742, USA}

\affiliation[f]{Carnegie Observatories, 
Pasadena, California 91101, USA}

\affiliation[\dagger]{Present address: High Energy Accelerator Research Organization (KEK), 
Tsukuba, Ibaraki 305-0801, Japan}

\emailAdd{xu@awa.tohoku.ac.jp}

\abstract{Geoneutrinos, electron antineutrinos produced by radioactive decays of heat-producing elements (HPEs) within the Earth, provide unique insights into Earth's interior and heat budget since their first detection in 2005 by KamLAND. Conventional geoneutrino detectors currently provide integrated global information and lack the capability to spatially resolve structures deep within the Earth. Here, we evaluate the ability of angular-sensitive geoneutrino detectors to distinguish between homogeneous and heterogeneous mantle models, focusing on Large Low Shear Velocity Provinces (LLSVPs). Our results show that LLSVPs enriched in Th and U yield a distinct flux of geoneutrinos with distinctive angular patterns. An oceanic site above the Pacific LLSVP is considered a particularly favorable detector location. The Ocean Bottom Detector (OBD) project aims to leverage this spatial resolving advantage by deploying a kiloton-scale liquid scintillator detector directly on the ocean floor, enabling unprecedented sensitivity for mantle geoneutrino detection. These findings demonstrate the critical role of combining geophysical and geochemical data to guide detector site selection, ultimately improving constraints on Earth's internal heat and the HPE distribution.}

\FullConference{19th International Conference on Topics in Astroparticle and Underground Physics (TAUP2025)\\
24–30 Aug 2025\\
Xichang, China\\}

\begin{document}
\maketitle

\section{Introduction}
According to geoscientific research, the Earth's interior is heterogeneous, featuring large-scale structures such as the Large Low Shear Velocity Provinces (LLSVPs). These are observed beneath the Pacific Ocean and Africa by seismic tomography as regions with anomalously slow S-wave velocities compared to the surrounding mantle \cite{Lay2006,Garnero2007}. 
Their origin remains uncertain, with hypotheses including enrichment in heat-producing elements such as uranium and thorium, among others, or accumulation of downwelling mantle heat \cite{McNamara2019}.

Geoneutrinos, emitted by beta decay of Earth's radioactive isotopes, provide a unique probe of radiogenic heat and elemental distribution \cite{KamLAND2005}. However, all existing geoneutrino detectors \cite{KamLAND2022,Borexino2020,SNO2025,JUNO2025} are situated in thick continental crusts that are rich in radioactive isotopes, and the resulting dominant crustal contribution masks the mantle signal, complicating the detection of mantle-origin geoneutrinos \cite{Sramek2013,Sramek2016}. To overcome this, the Ocean Bottom Detector (OBD) project proposes deploying a kiloton-scale liquid scintillator detector on Hawaii’s ocean floor \cite{Sakai2022,Watanabe2023}, where the crust is thin and contains fewer radioactive isotopes, making it an ideal site for mantle geoneutrino detection \cite{Rothschild1998}. In addition, previous geoneutrino measurements lack angular resolution, limiting spatial source identification. Advances such as gadolinium- or lithium-doped scintillators, gas-filled time projection chambers (TPCs), and segmented detectors hold strong potential to provide the directional sensitivity required to resolve mantle structures in future experiments \cite{Hochmuth2007,PROSPECT2025,Tanaka2014,Leyton2017,Duvall2024}.

In this study, we explore how angular resolution in future geoneutrino detectors can identify mantle heterogeneity with a focus on LLSVPs and constrain the distribution of heat-producing elements within them.

\section{Method}
\subsection{Nadir Angle Distribution of Geoneutrino Signals}
The geoneutrino signals \(\Phi(\bm{r})\) at detector position \(\bm{r}\) is given by Eq.~(\ref{flux_formula0}), where \(\varepsilon\) is the detection efficiency, \(N_A\) Avogadro's constant, \(\lambda_X\) the decay constant, and \(\mu_X\) the molar mass of isotope \(X\). The integral runs over the spatial region \(D(\theta)\) for nadir angle \(\theta\) and antineutrino energy \(E_\nu\), and includes the survival probability \(P_{\rm ee}\) as well as the rock density \(\rho(\bm{r}')\) and isotope abundance \(a_X(\bm{r}')\) at \(\bm{r}'\). The normalized spectrum \(dn_X/dE_\nu\) and IBD cross section \(\sigma_{\rm IBD}(E_\nu)\) are taken from \cite{Enomoto2005,Strumia2003}.

\begin{equation} \label{flux_formula0}
\Phi (\bm{r}) = \sum_{X \in \rm{HPE}} \frac{\varepsilon N_A \lambda_X}{\mu_X} \int_{D(\theta)} d\bm{r}' \int_{E_\nu} dE_\nu P_{\rm{ee}}(E_\nu, |\bm{r} - \bm{r}'|) \frac{\rho(\bm{r}') a_X(\bm{r}')}{4\pi |\bm{r} - \bm{r}'|^2} \frac{dn_X(E_\nu)}{dE_\nu} \sigma_{\rm{IBD}} (E_\nu)
\end{equation}

Among Earth's heat-producing elements, only antineutrinos from decay chains of \(^{238}\mathrm{U}\) and \(^{232}\mathrm{Th}\) are detectable via IBD \cite{Smirnov2019}. Thus, this study only considers geoneutrinos from these isotopes.

\subsection{Construction of Earth models}
\begin{figure}[htbp]
  \centering
  \includegraphics[width=1.0\textwidth]{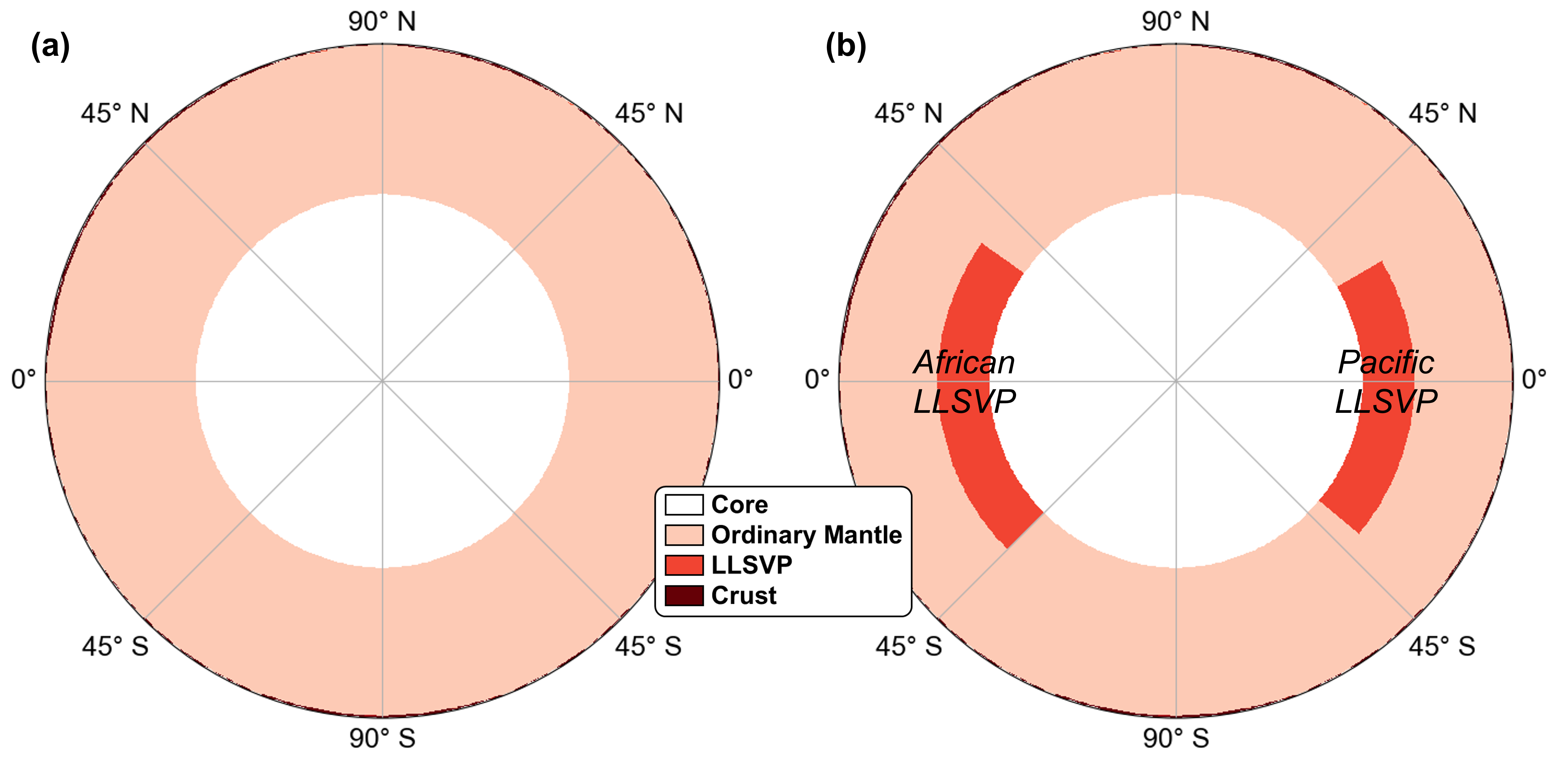}
  \caption{Simplified Earth models used in this study, showing the core, ordinary mantle, LLSVP, and crust partitions in a cross-section along the 180° meridian.
  \textbf{(a)} Homogeneous mantle model. \textbf{(b)} Heterogeneous mantle model with HPE-enriched LLSVP regions.}
  \label{fig:earth_models}
\end{figure}

In this study, we consider two simplified Earth models (Figure \ref{fig:earth_models}): (a) a homogeneous mantle model with uniform radionuclide abundances, and (b) a heterogeneous mantle model where LLSVP-like regions are enriched in radioactive elements.

To construct the models, we adopt the PREM density profile \cite{Dziewonski1981} for the mantle. Since uranium and thorium are lithophile elements largely absent from the core \cite{McDonough1995}, its contribution is neglected. Because geoneutrino flux depends inversely on the square of source-to-detector distance, accurate crust modeling is essential. Here, we use the CRUST1.0 model \cite{Laske2012}, a global $1^\circ \times 1^\circ$ dataset providing layer thickness, elevation, and density information.

Because the precise boundaries of the LLSVPs are not uniquely defined due to variations in seismic tomography methods and definitions \cite{Shephard2017}, we consider a simplified configuration in this study. Two LLSVP-like regions are modeled immediately above the core-mantle boundary (CMB): one beneath the Pacific Ocean with a radius of $35^\circ$ and one beneath Africa with a radius of $40^\circ$, both extending up to $1000\ \mathrm{km}$ in height and located at nearly antipodal positions.

Chemical compositions for both the crust and the mantle follow \cite{Huang2013,Arevalo2010}. In the homogeneous model, the mantle is assigned the globally averaged mantle HPE abundances from \cite{Huang2013}. For the heterogeneous model, we conserve the HPE inventory attributed to the enriched mantle in \cite{Huang2013} but confine it specifically to the LLSVP-like regions. Note that the enriched mantle domain in \cite{Huang2013} is spatially larger than the LLSVPs modeled in this study. Consequently, the remaining ordinary mantle is assigned depleted mantle abundances based on Mid-Ocean Ridge Basalt (MORB) data \cite{Huang2013,Arevalo2010} (see Table~\ref{tab:HPEabundance}).

\begin{table}[htbp]
\centering
\caption{HPE concentrations in mantle regions of the two Earth models}
\begin{tabular}{lcccc}
\hline
\multirow{2}{*}{Region} 
  & \multicolumn{2}{c}{(a) Homogeneous mantle model} 
  & \multicolumn{2}{c}{(b) Heterogeneous mantle model} \\
\cline{2-3} \cline{4-5}
 & U [ppb] & Th [ppb] & U [ppb] & Th [ppb] \\
\hline
Ordinary mantle    & 12.7 & 47.2   & 8.0 & 22.0 \\
LLSVP-like regions & -    & -      & 93.7 & 483.4 \\
\hline
\end{tabular}
\label{tab:HPEabundance}
\end{table}

\subsection{Application of angular resolution}
Accurate evaluation of angular resolution requires detailed simulations, which are beyond the scope of this study since the kiloton-scale OBD is still under design and detailed studies remain limited.  
Instead, we assess the utility of angular sensitivity by considering a range of angular resolutions and comparing the resulting smeared nadir angle distributions for the two Earth models. 
From this, we quantify how model separability depends on angular resolution and event statistics, and whether directional information can meaningfully constrain LLSVP chemical enrichment.

\section{Result \& Discussion}
\subsection{Expected detection result at the Pacific Ocean}
Table \ref{tab:geoneutrino_flux} summarizes the geoneutrino signal rates at a representative Pacific location (180°E, 5°S) under two different Earth models, expressed in Terrestrial Neutrino Units (TNU).
1 TNU corresponds to 1 geoneutrino event detectable per 
$10^{32}$ target protons per year, 
providing a standard measure for geoneutrino detection rates \cite{Fiorentini2003}.
Note that this coordinate is not the only candidate location for the OBD project. In this study, it is used purely for computational convenience, as it lies directly above the Pacific LLSVP as constructed in our model.

\begin{table}[htbp]
\centering
\caption{Geoneutrino signal rate at the Pacific Ocean under different Earth models (in TNU)}
\begin{tabular}{lcc}
\hline
Component & (a) Homogeneous mantle model & (b) Heterogeneous mantle model \\
\hline
Ordinary mantle    & 10.1  & 5.8  \\
LLSVP-like regions  & -  & 6.2  \\
Continental crust  & 2.3  & 2.3  \\
Oceanic crust     & 0.9  & 0.9  \\
\hline
Total signal rate   & 13.3 & 15.2 \\
\hline
\end{tabular}
\label{tab:geoneutrino_flux}
\end{table}

\begin{figure}[htbp]
  \centering
  \includegraphics[width=1.0\textwidth]{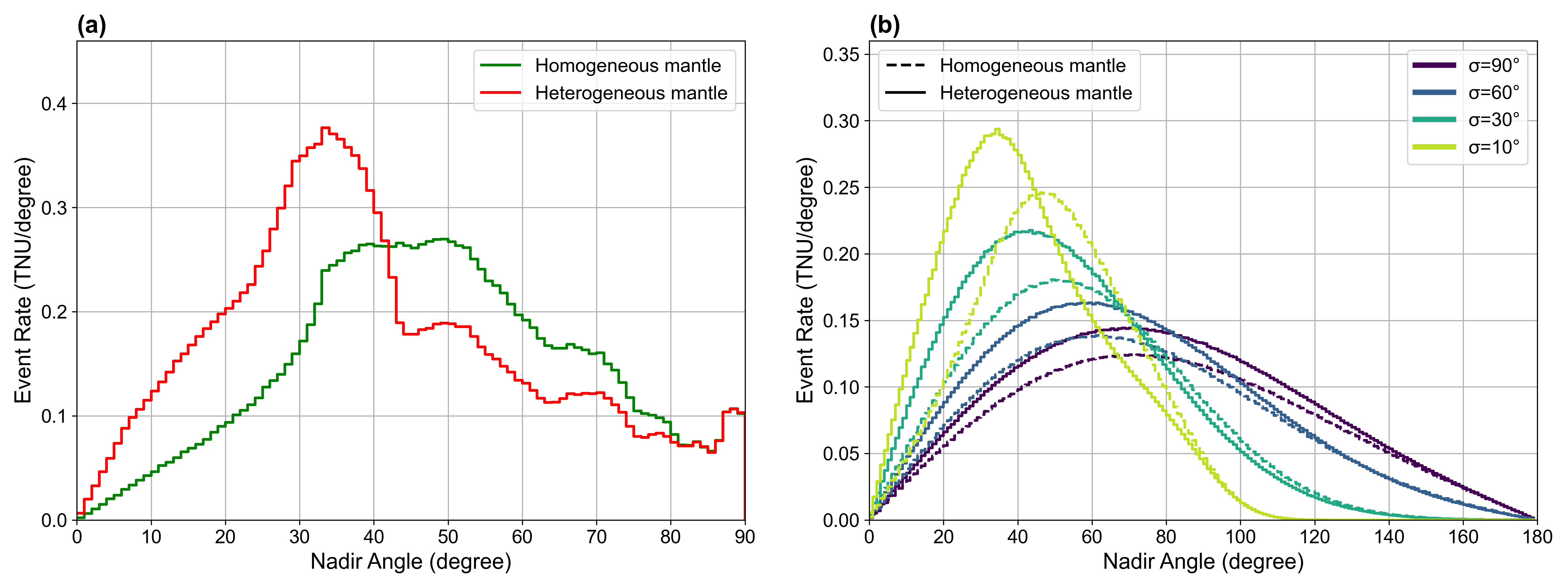}
    \caption{\textbf{(a)} Nadir angle distributions of geoneutrino signals (in TNU); \textbf{(b)} Expected detection result after applying angular resolution $\sigma=(10, 30, 60, 90)^\circ$. Note that the $x$-axis ranges differ between the two subplots.}
  \label{fig:geoneutrino_flux}
\end{figure}

Figure \ref{fig:geoneutrino_flux} compares the nadir angle distributions of geoneutrino signals from two Earth models. Panel (a) shows the raw distributions, while panel (b) illustrates the expected detection result after accounting for detector angular resolution. 
The differences between the two models reveal the effect of mantle heterogeneity on the detectable signals, indicating that directional measurements can help identify LLSVPs and constrain the heat-producing element concentrations within them. 

\subsection{Required angular resolution and experimental exposure}

Figure \ref{fig:geoneutrino_flux}(b) indicates that improved angular resolution increases the ability to discriminate between the two Earth models and to constrain HPE abundances in the LLSVPs. To offer insight relevant to future OBD designs, we therefore calculate the combinations of angular resolution and exposure required to distinguish two Earth models at 1–5\,$\sigma$ significance. Exposure is defined as the product of detector live time and the effective number of target protons. The significance is evaluated based on a standard Figure of Merit (FoM), defined as the difference between the mean values of the two model predictions divided by the combined standard deviation.

\begin{figure}[htbp]
  \centering
  \includegraphics[width=0.71\textwidth]{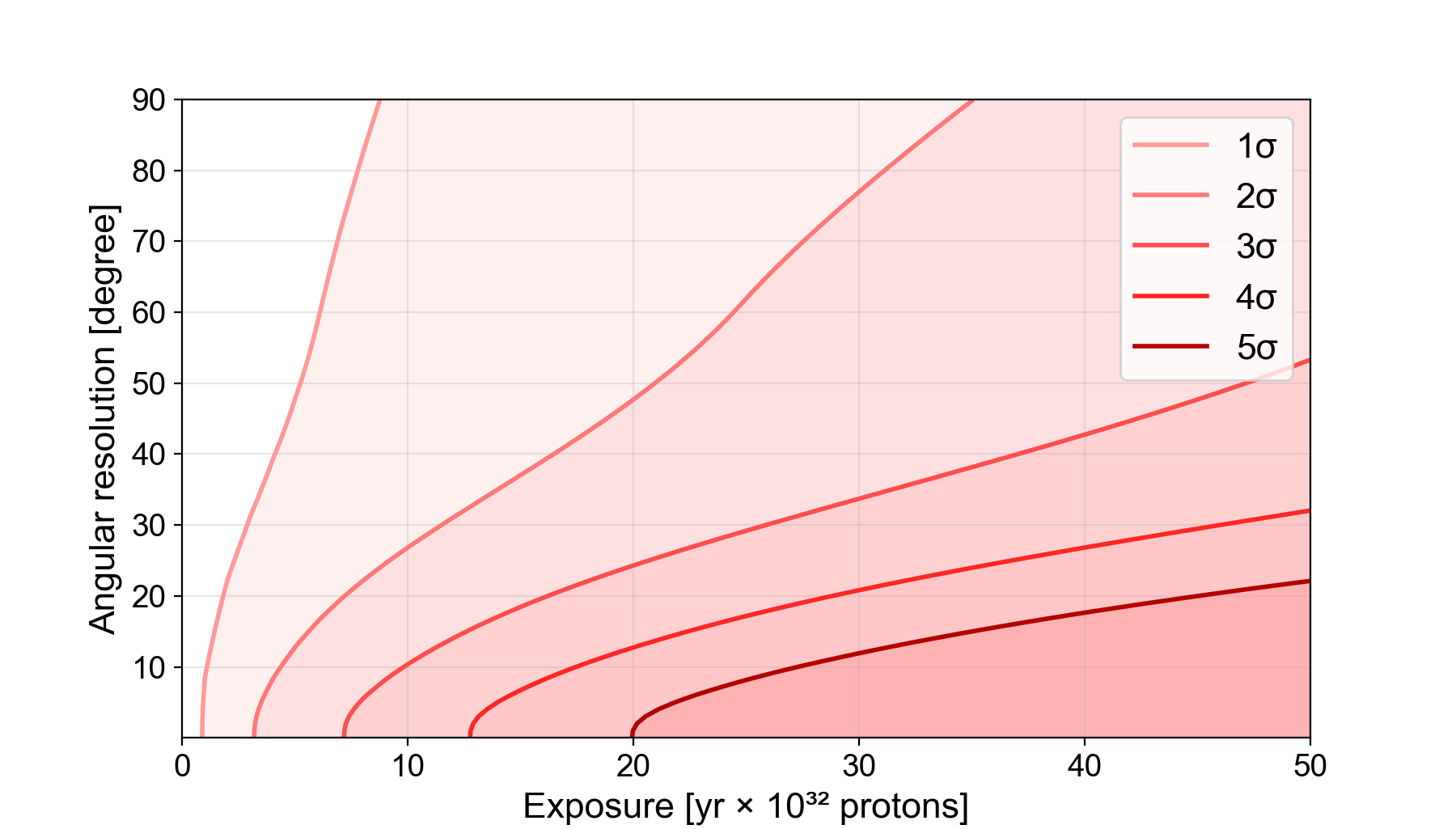}
  \caption{Combinations of angular resolution (Gaussian $\sigma$ in degrees) and exposure (effective proton-years) required to distinguish the homogeneous and the heterogeneous mantle models at 1–5\,$\sigma$ significance.}
  \label{fig:ARvsE}
\end{figure}

The resulting combinations of angular resolution and exposure that achieve 1–5\,$\sigma$ significance are plotted in Figure \ref{fig:ARvsE}. Larger exposure increases the detected event count, thereby reducing statistical uncertainty and improving the achievable confidence level.

We note that this assessment neglects background signals, detection efficiency and systematic uncertainties. Thus, the requirements shown here should be regarded as optimistic lower bounds. A more realistic evaluation will be conducted in future work, potentially increasing exposure and/or angular resolution demands.

\section{Conclusion}

Our study demonstrates that angular-sensitive geoneutrino detection can effectively distinguish between homogeneous and heterogeneous mantle models, particularly revealing the presence and chemical characteristics of Large Low Shear Velocity Provinces (LLSVPs).
Considering that oceanic regions offer the most favorable conditions for observing mantle-origin geoneutrinos, the planned Ocean Bottom Detector (OBD) project represents a significant advance by deploying a detector directly on the seafloor, thereby enabling unprecedented sensitivity through minimized crustal background interference.

\section*{Acknowledgment}
This work was supported by JSPS KAKENHI Grants No. 24K00653. This work was performed under the auspices of the U.S. Department of Energy by Lawrence Livermore National Laboratory under Contract DE-AC52-07NA27344. LLNL-PROC-2013747.


\begin{thebibliography}{99}

\bibitem{Lay2006}
T.~Lay \textit{et al.},
\emph{A post-perovskite lens and D'' heat flux beneath the central Pacific},
\href{https://www.science.org/doi/10.1126/science.1133280}{\emph{Science} \textbf{314} (2006) 1272-1276.}

\bibitem{Garnero2007}
E.~J.~Garnero and A.~K.~McNamara,
\emph{Structure and dynamics of Earth's lower mantle},
\href{https://www.science.org/doi/10.1126/science.1148028}{\emph{Science} \textbf{320} (2008) 626-628.}

\bibitem{McNamara2019}
A.~K.~McNamara,
\emph{A review of large low shear velocity provinces and ultra low velocity zones},
\href{https://doi.org/10.1016/j.tecto.2018.04.015}{\emph{Tectonophysics} \textbf{760} (2019) 199-220.}

\bibitem{KamLAND2005}
T.~Araki \textit{et al.} (KamLAND Collaboration),
\emph{Experimental investigation of geologically produced antineutrinos with KamLAND},
\href{https://doi.org/10.1038/nature03980}{\emph{Nature} \textbf{436} (2005) 499-503.}

\bibitem{KamLAND2022}
S.~Abe \textit{et al.} (KamLAND Collaboration),
\emph{Abundances of uranium and thorium elements in Earth estimated by geoneutrino spectroscopy},
\href{https://doi.org/10.1029/2022GL099566}{\emph{Geophys. Res. Lett.} \textbf{49} (2022) e2022GL099566.}

\bibitem{Borexino2020}
M.~Agostini \textit{et al.} (Borexino Collaboration),
\emph{Comprehensive geoneutrino analysis with Borexino},
\href{https://journals.aps.org/prd/abstract/10.1103/PhysRevD.101.012009}{\emph{Phys. Rev. D} \textbf{101} (2020) 012009.}

\bibitem{SNO2025}
M.~Abreu \textit{et al.} (SNO+ Collaboration),
\emph{Measurement of Reactor Antineutrino Oscillation at SNO+},
\href{https://link.aps.org/doi/10.1103/gypt-lc9v}{\emph{Phys. Rev. Lett.} \textbf{135} (2025) 121801.}

\bibitem{JUNO2025}
A.~Abusleme \textit{et al.} (JUNO Collaboration),
\emph{First measurement of reactor neutrino oscillations at JUNO},
\href{https://arxiv.org/abs/2511.14593}{[{\tt arXiv:2511.14593}]}.

\bibitem{Sramek2013}
O.~Šrámek \textit{et al.},
\emph{Geophysical and geochemical constraints on geoneutrino fluxes from Earth's mantle},
\href{https://doi.org/10.1016/j.epsl.2012.11.001}{\emph{Earth Planet. Sci. Lett.} \textbf{361} (2013) 356-366.}

\bibitem{Sramek2016}
O.~Šrámek \textit{et al.},
\emph{Revealing the Earth's mantle from the tallest mountains using the Jinping Neutrino Experiment},
\href{https://doi.org/10.1038/srep33034}{\emph{Sci. Rep.} \textbf{6} (2016) 33034.}

\bibitem{Sakai2022}
T.~Sakai \textit{et al.},
\emph{Study of Ocean Bottom Detector for observation of geo-neutrino from the mantle},
\href{https://iopscience.iop.org/article/10.1088/1742-6596/2156/1/012144}{\emph{J. Phys. Conf. Ser.} \textbf{2156} (2022) 012144.}

\bibitem{Watanabe2023}
H.~Watanabe \textit{et al.},
\emph{Ocean Bottom Detector: frontier of technology for understanding the mantle by geoneutrinos},
\href{https://ieeexplore.ieee.org/abstract/document/10103417/}{\textit{2023 IEEE Underwater Technology} (2023).}

\bibitem{Rothschild1998}
C.~G.~Rothschild \textit{et al.},
\emph{Antineutrino geophysics with liquid scintillator detectors},
\href{https://agupubs.onlinelibrary.wiley.com/doi/abs/10.1029/98GL50667}{\emph{Geophys. Res. Lett.} \textbf{25} (1998) 1083-1086.}

\bibitem{Hochmuth2007}
K.~A.~Hochmuth \textit{et al.},
\emph{Exploiting the directional sensitivity of the Double Chooz near detector},
\href{https://doi.org/10.1103/PhysRevD.76.073001}{\emph{Phys. Rev. D} \textbf{76} (2007) 073001.}

\bibitem{PROSPECT2025}
M.~Andriamirado \textit{et al.} (PROSPECT Collaboration),
\emph{Reactor antineutrino directionality measurement with the PROSPECT-I detector},
\href{https://journals.aps.org/prd/abstract/10.1103/PhysRevD.111.032014}{\textit{Phys. Rev. D} \textbf{111} (2025) 032014.}

\bibitem{Tanaka2014}
H.~K.~M.~Tanaka and H.~Watanabe,
\emph{$^{6}$Li-loaded directionally sensitive anti-neutrino detector for possible geo-neutrinographic imaging applications},
\href{https://doi.org/10.1038/srep04708}{\emph{Sci. Rep.} \textbf{4} (2014) 4708.}

\bibitem{Leyton2017}
M.~Leyton \textit{et al.},
\emph{Exploring the hidden interior of the Earth with directional neutrino measurements},
\href{https://www.nature.com/articles/ncomms15989}{\emph{Nat. Commun.} \textbf{8} (2017) 15989.}

\bibitem{Duvall2024}
M.~J.~Duvall \textit{et al.},
\emph{Directional response of several geometries for reactor-neutrino detectors},
\href{https://doi.org/10.1103/PhysRevApplied.22.054030}{\emph{Phys. Rev. Appl.} \textbf{22} (2024) 054030.}

\bibitem{Enomoto2005}
S.~Enomoto,
\emph{Neutrino geophysics and observation of geo-neutrinos at KamLAND},
Ph.D. Thesis, Tohoku University, (2005).

\bibitem{Strumia2003}
A.~Strumia and F.~Vissani,
\emph{Precise quasielastic neutrino/nucleon cross-section},
\href{https://doi.org/10.1016/S0370-2693(03)00616-6}{\emph{Phys. Lett. B} \textbf{564} (2013) 42-54.}

\bibitem{Smirnov2019}
O.~Smirnov,
\emph{Experimental aspects of geoneutrino detection: Status and perspectives},
\href{https://www.sciencedirect.com/science/article/pii/S014664101930047X}{\emph{Prog. Part. Nucl. Phys.} \textbf{109} (2019) 103712.}

\bibitem{Dziewonski1981}
A.~M.~Dziewonski and D.~L.~Anderson,
\emph{Preliminary reference Earth model},
\href{https://doi.org/10.1016/0031-9201(81)90046-7}{\emph{Phys. Earth Planet. Inter.} \textbf{25} (1981) 297-356.}

\bibitem{McDonough1995}
W.~F.~McDonough and S.-s.~Sun,
\emph{The composition of the Earth},
\href{https://doi.org/10.1016/0009-2541(94)00140-4}{\emph{Chem. Geol.} \textbf{120} (1995) 223-253.}

\bibitem{Laske2012}
G.~Laske \textit{et al.},
\emph{CRUST1.0: An updated global model of Earth's crust},
\href{https://meetingorganizer.copernicus.org/EGU2012/EGU2012-3743-1.pdf}{\emph{Geophys. Res. Abstr.} \textbf{14} (2012) 743.}

\bibitem{Shephard2017}
G.~E.~Shephard \textit{et al.},
\emph{On the consistency of seismically imaged lower mantle slabs},
\href{https://doi.org/10.1038/s41598-017-11039-w}{\emph{Sci. Rep.} \textbf{7} (2017) 10976.}

\bibitem{Huang2013}
Y.~Huang \textit{et al.},
\emph{A reference Earth model for the heat-producing elements and associated geoneutrino flux},
\href{https://doi.org/10.1002/ggge.20129}{\emph{Geochem. Geophys. Geosyst.} \textbf{14} (2013) 2003-2029.}

\bibitem{Arevalo2010}
R.~Arevalo~Jr. and W.~F.~McDonough,
\emph{Chemical variations and regional diversity observed in MORB},
\href{https://www.sciencedirect.com/science/article/pii/S000925410900480X}{\emph{Chem. Geol.} \textbf{271} (2010) 70-85.}

\bibitem{Fiorentini2003}
G.~Fiorentini \textit{et al.},
\emph{KamLAND, terrestrial heat sources and neutrino oscillations},
\href{https://doi.org/10.1016/S0370-2693(03)00240-5}{\emph{Phys. Lett. B} \textbf{558} (2003) 15-21.}


\end{thebibliography}
\end{document}